\documentclass[fleqn]{llncs}

\usepackage{amsmath}
\usepackage{amsfonts}
\usepackage{exscale}
\usepackage{calc}
\usepackage{epsfig}
\usepackage{url}
\usepackage{notations}

\begin{document}
\newcommand{\eqv}{=\!= \;}

\newcommand{\lar}{\Longleftarrow \;}

\newcommand{\rar}{\Longrightarrow \;}

\newcommand{\br}{\displaybreak[0] \\}

\newcommand{\tensor}{\otimes}

\newcommand{\xor}{\oplus}

\newcommand{\entrylabel}[1]{%
                 \mbox{\textbf{#1:}}\hfil}
\newenvironment{entry}%
  {\begin{list}{}{%
    \renewcommand{\makelabel}{\entrylabel}%
    \setlength{\labelwidth}{35pt}%
    \setlength{\leftmargin}%
                 {\labelwidth+\labelsep}}}%
  {\end{list}}

\newcommand{\ket}[1]{|#1\rangle}
\newcommand{\bra}[1]{\langle#1|}
\newcommand{\braket}[2]{\langle #1 | #2 \rangle}

\newcommand{\ketb}[1]{|\textbf{#1}\rangle}
\newcommand{\brab}[1]{\langle\textbf{#1}|}
\newcommand{\braketb}[2]{\langle \textbf{#1} | \textbf{#2} \rangle}

\frontmatter
\pagestyle{headings}
\addtocmark{Quantum Predicative Programming}

\mainmatter
\title{Quantum Predicative Programming}
\titlerunning{Quantum Predicative Programming}
\author{Anya Tafliovich \and E.C.R. Hehner}
\authorrunning{A. Tafliovich and E.C.R. Hehner}
\tocauthor{Anya Tafliovich (University of Toronto),
           E.C.R. Hehner (University of Toronto)}
\institute{University of Toronto}

\maketitle

\begin{abstract}
  The subject of this work is quantum predicative programming --- the
  study of developing of programs intended for execution on a quantum
  computer.  We look at programming in the context of formal methods
  of program development, or programming methodology.  Our work is
  based on probabilistic predicative programming, a recent
  generalisation of the well-established predicative programming. It
  supports the style of program development in which each programming
  step is proven correct as it is made.  We inherit the advantages of
  the theory, such as its generality, simple treatment of recursive
  programs, time and space complexity, and communication. Our theory
  of quantum programming provides tools to write both classical and
  quantum specifications, develop quantum programs that implement
  these specifications, and reason about their comparative time and
  space complexity all in the same framework.
\end{abstract}


\section{Introduction}
\label{sec:introduction}

Modern physics is dominated by concepts of quantum mechanics. Today,
over seventy years after its recognition by the scientific community,
quantum mechanics provides the most accurate known description of
nature's behaviour. Surprisingly, the idea of using the quantum
mechanical nature of the world to perform computational tasks is very
new, less than thirty years old. Quantum computation and quantum
information is the study of information processing and communication
accomplished with quantum mechanical systems. In recent years the
field has grown immensely. Scientists from various fields of computer
science have discovered that thinking physically about computation
yields new and exciting results in computation and communication.
There has been extensive research in the areas of quantum algorithms,
quantum communication and information, quantum cryptography, quantum
error-correction, adiabatic computation, measurement-based quantum
computation, theoretical quantum optics, and the very new quantum game
theory. Experimental quantum information and communication has also
been a fruitful field. Experimental quantum optics, ion traps, solid
state implementations and nuclear magnetic resonance all add to the
experimental successes of quantum computation.

The subject of this work is quantum programming --- the study of
developing programs intended for execution on a quantum computer.  We
assume a model of a quantum computer proposed by Knill~\cite{knill96}:
a classical computer with access to a quantum device that is capable
of storing quantum bits, performing certain operations and
measurements on these bits, and reporting the results of the
measurements.

We look at programming in the context of formal methods of program
development, or programming methodology. This is the field of computer
science concerned with applications of mathematics and logic to
software engineering tasks. In particular, the formal methods provide
tools to formally express software specifications, prove correctness
of implementations, and reason about various properties of
specifications (e.g. implementability) and implementations (e.g. time
and space complexity). Today formal methods are successfully employed
in all stages of software development, such as requirements
elicitation and analysis, software design, and software
implementation.

In this work the theory of quantum programming is based on
probabilistic predicative programming, a recent generalisation of the
well-established predicative
programming~\cite{hehner04:_book_2ed,hehner04:_ppp}, which we deem to
be the simplest and the most elegant programming theory known today.
It supports the style of program development in which each programming
step is proven correct as it is made.  We inherit the advantages of
the theory, such as its generality, simple treatment of recursive
programs, and time and space complexity. Our theory of quantum
programming provides tools to write both classical and quantum
specifications, develop quantum programs that implement these
specifications, and reason about their comparative time and space
complexity all in the same framework.

The rest of this work is organised as follows.  Section~\ref{sec:qc}
is the introduction to quantum computation. It assumes that the reader
has some basic knowledge of linear algebra and no knowledge of quantum
computing.  Section~\ref{sec:ppp} contains the introduction to
probabilistic predicative programming. The reader is assumed to have
some background in logic, but no background in programming theory is
necessary. The contribution of this work is section~\ref{sec:qpp}
which defines the quantum system, introduces programming with the
quantum system, and several well-known problems, their classical and
quantum solutions, and their formal comparative time complexity
analyses.  Section~\ref{sec:conclusion} states conclusions and
outlines directions for future research.

\subsection{Related work}
\label{sec:relwork}

Traditionally, quantum computation is presented in terms of quantum
circuits. Recently, there has been an attempt to depart from this
convention for the same reason that classical computation is generally
not presented in terms of classical circuits. As we develop more
complex quantum algorithms, we will need ways to express higher-level
concepts with control structures in a readable fashion.

In 2000 {\"O}mer~\cite{omer00} introduced the first quantum
programming language QCL. Following his work, Bettelli \emph{et.
  al.} developed a quantum programming language with syntax based on
C++. These two works did not involve any verification techniques.

Sanders and Zuliani in~\cite{zuliani00} introduced a quantum language
qGCL, which is an extension of pGCL~\cite{morgan99}, which in turn
generalises Dijkstra's guarded-command language to include
probabilism. Zuliani later extends this attempt at formal program
development and verification in~\cite{zuliani04}, which discusses
treatment of non-determinism in quantum programs, and
in~\cite{zuliani05}, where the attempt is made to build on Aharonov's
work to reason about mixed states computations. Zuliani also provides
tools to approach the task of compiling quantum programs
in~\cite{zuliani05:journal}. 

A large amount of work in the area was performed in the past two
years. In~\cite{adao05},~\cite{lalire04}, and~\cite{jorrand04}
process algebraic approaches were explored. Tools developed in the
field of category theory were successfully employed
by~\cite{abramsky04},~\cite{abramsky_coecke04},~\cite{abramsky_duncan04},
~\cite{coecke04_logic},~\cite{selinger04}, and others to reason about
quantum computation.  In~\cite{arrighi04} and~\cite{arrighi05} a
functional language with semantics in a form of a term rewrite system
is introduced and a notion of linearity and how it pertains to quantum
systems are examined. A functional language QML with design guided by
its categorical semantics is defined in~\cite{altenkirch05_fqpl}.
Following on this work,~\cite{altenkirch05} provides a sound and
complete equational theory for QML. Weakest preconditions appropriate
for quantum computation are introduced in~\cite{d'hondt04}. This work
is interesting, in part, because it diverts from the standard approach
of reducing quantum computation to probabilistic one. It also provides
semantics for the language of~\cite{selinger04}. Other interesting
work by the same authors include reasoning about knowledge in quantum
systems (\cite{d'hondt05}) and developing a formal model for distributed
measurement-based quantum computation (\cite{danos05}). A similar work
is introduced in~\cite{gay05}, where a language CQP for modelling
communication in quantum systems is defined. The latter approaches have
an advantage over process algebraic approaches mentioned earlier in
that they explicitly allow a quantum state to be transmitted between
processes. Building of the work
of~\cite{selinger04s},~\cite{valiron04} defines a higher order quantum
programming language based on a linear typed lambda calculus, which is
similar to the work of~\cite{tonder04}.

\subsection{Our contribution}
\label{sec:contribution}

Our approach to quantum programming amenable to formal analysis is
very different from almost all of those described above. Work
of~\cite{zuliani00},~\cite{zuliani04},~\cite{zuliani05} is the only
one which is similar to our work. The contribution of this paper is
twofold.  Firstly, by building our theory on that
in~\cite{hehner04:_ppp}, we inherit the advantages it offers. The
definitions of specification and program are simpler: a specification
is a boolean (or probabilistic) expression and a program is a
specification. The treatment of recursion is simple: there is no need
for additional semantics of loops. The treatment of termination simply
follows from the introduction of a time variable; if the final value
of the time variable is $\infty$, then the program is a
non-terminating one.  Correctness and time and space complexity are
proved in the same fashion; moreover, after proving them separately,
we naturally obtain the conjunction. Secondly, the way Probabilistic
Predicative Programming is extended to Quantum Predicative Programming
is simple and intuitive. The use of Dirac-like notation makes it easy
to write down specifications and develop algorithms. The treatment of
computation with mixed states does not require any additional
mechanisms. Quantum Predicative Programming fully preserves
Predicative Programming's treatment of parallel programs and
communication, which provides for a natural extension to reason about
quantum communication protocols, such as BB84 (\cite{bennet84}),
distributed quantum algorithms, such as distributed Shor's algorithm
(\cite{yimsiriwattana04}), as well as their time, space, and
entanglement complexity.


\section{Preliminaries}
\label{sec:preliminaries}


\subsection {Quantum Computation}
\label{sec:qc}

In this section we introduce the basic concepts of quantum mechanics,
as they pertain to the quantum systems that we will consider for quantum
computation. The discussion of the underlying physical processes,
spin-$\frac{1}{2}$-particles, etc.~is not of our interest. We are
concerned with the model for quantum computation only.  A reader not
familiar with quantum computing can consult~\cite{nielsen00} for a
comprehensive introduction to the field.

The \emph{Dirac notation}, invented by Paul Dirac, is often used in
quantum mechanics. In this notation a vector $v$ (a column vector by
convention) is written inside a \emph{ket}: $\ket{v}$.  The dual
vector of $\ket{v}$ is $\bra{v}$, written inside a \emph{bra}. The
inner products are \emph{bra-kets} $\braket{v}{w}$.  For
$n$-dimensional vectors $\ket{u}$ and $\ket{v}$ and $m$-dimensional
vector $\ket{w}$, the value of the inner product $\braket{u}{v}$ is a
scalar and the outer product operator $\ket{v}\bra{w}$ corresponds to
an $m$ by $n$ matrix.  The Dirac notation clearly distinguishes
vectors from operators and scalars, and makes it possible to write
operators directly as combinations of bras and kets.

In quantum mechanics, the vector spaces of interest are the Hilbert
spaces of dimension $2^n$ for some $n \in \mathbb{N}$.  A convenient
orthonormal basis is what is called a \emph{computational basis}, in
which we label $2^n$ basis vectors using binary strings of length $n$
as follows: if $s$ is an $n$-bit string which corresponds to the
number $x_s$, then $\ket{s}$ is a $2^n$-bit (column) vector with $1$
in position $x_s$ and $0$ everywhere else. The tensor product $\ket{i}
\tensor \ket{j}$ can be written simply as $\ket{ij}$.  An arbitrary
vector in a Hilbert space can be written as a weighted sum of the
computational basis vectors.

\begin{description}
\item[Postulate 1 (state space)] Associated to any isolated physical
  system is a Hilbert space, known as the \emph{state space} of the
  system. The system is completely described by its \emph{state
    vector}, which is a unit vector in the system's state space.
\end{description}
\begin{description}
\item [Postulate 2 (evolution)] The evolution of a closed quantum
  system is described by a \emph{unitary transformation}.
\end{description}
\begin{description}
\item [Postulate 3 (measurement)] Quantum measurements are described
  by a collection $\{M_m\}$ of \emph{measurement operators}, which act
  on the state space of the system being measured. The index $m$
  refers to the possible measurement outcomes. If the state of the
  system immediately prior to the measurement is described by a vector
  $\ket{\psi}$, then the probability of obtaining result $m$ is
  $\braket{\psi}{M_m^{\dagger} M_m |\psi}$, in which case the state of
  the system immediately after the measurement is described by the
  vector $\frac{M_m \ket{\psi}}{\sqrt{\braket{\psi}{M_m^{\dagger} M_m
        |\psi}}}$. The measurement operators satisfy the
  \emph{completeness equation} $\sum m \cdot M_m^{\dagger} M_m = I$.
\end{description}
An important special class of measurements is \emph{projective
  measurements}, which are equivalent to general measurements provided
that we also have the ability to perform unitary transformations.

A projective measurement is described by an \emph{observable} $M$,
which is a Hermitian operator on the state space of the system being
measured. This observable has a spectral decomposition $M=\sum m \cdot
\lambda_m \times P_m$, where $P_m$ is the projector onto the
eigenspace of $M$ with eigenvalue $\lambda_m$, which corresponds to
the outcome of the measurement.  The probability of measuring $m$ is
$\braket{\psi}{P_m | \psi}$, in which case immediately after the
measurement the system is found in the state $\frac{P_m
  \ket{\psi}}{\sqrt{\braket{\psi}{P_m | \psi}}}$.

Given an orthonormal basis $\ket{v_m}$, $0 \leq m < 2^n$, measurement
with respect to this basis is the corresponding projective measurement
given by the observable $M = \sum m \cdot \lambda_m \times P_m$, where the
projectors are $P_m = \ket{v_m}\bra{v_m}$.

Measurement with respect to the computational basis is the simplest
and the most commonly used class of measurements. In terms of the
basis $\ket{m}$, $0 \leq m < 2^n$, the projectors are $P_m =
\ket{m}\bra{m}$ and $\braket{\psi}{P_m | \psi} = |\psi_m|^2$. The
state of the system immediately after measuring $m$ is $\ket{m}$.

In the case of a single qubit, for example, measurement of the state
$\alpha \times \ket{0} + \beta \times \ket {1}$ results in the outcome
$0$ with probability $|\alpha|^2$ and outcome $1$ with probability
$|\beta|^2$.  The state of the system immediately after the
measurement is $\ket{0}$ or $\ket{1}$, respectively.

Suppose the result of the measurement is ignored and we continue the
computation. In this case the system is said to be in a \emph{mixed
  state}. A mixed state is not the actual physical state of the
system. Rather it describes our knowledge of the state the system is
in.  In the above example, the mixed state is expressed by the
equation $\ket{\psi} = |\alpha|^2 \times \{\ket{0}\} + |\beta|^2
\times \{\ket{1}\}$. The equation is meant to say that $\ket{\psi}$ is
$\ket{0}$ with probability $|\alpha|^2$ and it is $\ket{1}$ with
probability $|\beta|^2$. An application of operation $U$ to the mixed
state results in another mixed state, $U(|\alpha|^2 \times \{\ket{0}\}
+ |\beta|^2 \times \{\ket{1}\}) = |\alpha|^2 \times \{U\ket{0}\} +
|\beta|^2 \times \{U\ket{1}\}$.

\begin{description}
\item [Postulate 4 (composite systems)] The state space of a composite
  physical system is the tensor product of the state spaces of the
  component systems. If we have systems numbered $0$ up to and
  excluding $n$, and each system $i$, $0 \leq i < n$, is prepared in the state
  $\ket{\psi_i}$, then the joint state of the composite system is
  $\ket{\psi_0} \tensor \ket{\psi_1} \tensor \ldots \tensor
  \ket{\psi_{n-1}}$.
\end{description}
While we can always describe a composite system given descriptions of
the component systems, the reverse is not true. Indeed, given a state
vector that describes a composite system, it may not be possible to
factor it to obtain the state vectors of the component systems. A
well-known example is the state $\ket{\psi} = \ket{00}/\sqrt{2} +
\ket{11}/\sqrt{2}$. Such state is called an \emph{entangled} state.


\subsection {Probabilistic Predicative Programming}
\label{sec:ppp}

This section introduces the programming theory of our choice, on which
our work on quantum programming is based --- probabilistic predicative
programming. We briefly introduce parts of the theory necessary for
understanding section~\ref{sec:qpp} of this work. For a course in
predicative programming the reader is referred
to~\cite{hehner04:_book_2ed}. Introduction to probabilistic
predicative programming can be found in~\cite{hehner04:_ppp}.

\subsubsection{Predicative programming}
\label{sec:pp}

In predicative programing a specification is a boolean expression. The
variables in a specification represent the quantities of interest,
such as prestate (inputs), poststate (outputs), and computation time
and space. We use primed variables to describe outputs and unprimed
variables to describe inputs.  For example, specification $x' = x + 1$
in one integer variable $x$ states that the final value of $x$ is its
initial value plus $1$.  A computation \emph{satisfies} a
specification if, given a prestate, it produces a poststate, such that
the pair makes the specification true. A specification is
\emph{implementable} if for each input state there is at least one
output state that satisfies the specification.

We use standard logical notation for writing specifications: $\wedge$
(conjunction), $\vee$ (disjunction), $\Rightarrow$ (logical
implication), $=$ (equality, boolean equivalence), $\neq$
(non-equality, non-equivalence), and \textbf{if then else}. $\eqv$ and
$\Longrightarrow$ are the same as $=$ and $\Rightarrow$, but with
lower precedence. We use standard mathematical notation, such as $ +
\, - \, * \, / \, mod $.  We use lowercase letters for variables of
interest and uppercase letters for specifications.

In addition to the above, we use the following notations: $\sigma$
(prestate), $\sigma'$ (poststate), $ok$ ($\sigma'=\sigma$), and $x:=e$
($x'=e \wedge y'=y \wedge \hdots$). $ok$ specifies that the values of
all variables are unchanged. In the assignment $x:=e$, $x$ is a state
variable (unprimed) and $e$ is an expression (in unprimed variables)
in the domain of $x$.

If $R$ and $S$ are specifications in variables $x, y, \hdots \;$,
$R''$ is obtained from $R$ by substituting all occurrences of primed
variables $x', y', \hdots$ with double-primed variables $x'', y'',
\hdots \;$, and $S''$ is obtained from $S$ by substituting all
occurrences of unprimed variables $x, y, \hdots$ with double-primed
variables $x'', y'', \hdots \;$, then the \emph{sequential
  composition} of $R$ and $S$ is defined by 

$$R;S \eqv \exists x'', y'', \hdots \cdot R'' \wedge S''$$.

Various laws can be proven about sequential composition. One of the
most important ones is the substitution law, which states that for any
expression $e$ of the prestate, state variable $x$, and specification
$P$,

$$x:=e; P \eqv (\text{for } x \text{ substitute } e \text{ in } P)$$

Specification $S$ \emph{is refined by} specification $P$ if and only
if $S$ is satisfied whenever $P$ is satisfied: 

$$\forall \sigma, \sigma' \cdot S \Leftarrow P$$

Specifications $S$ and $P$ are equal if and only if they are satisfied
simultaneously: 

$$\forall \sigma, \sigma' \cdot S = P$$

Given a specification, we are allowed to implement an equivalent
specification or a stronger one.

Informally, a \emph{bunch} is a collection of objects. It is different
from a set, which is a collection of objects in a package. Bunches are
simpler than sets; they don't have a nesting structure.
See~\cite{hehner04:_ppp} for an introduction to bunch theory. A bunch
of one element is the element itself.  We use upper-case to denote
arbitrary bunches and lower-case to denote elements (an element is the
same as a bunch of one element). $A, B$ denotes the union of bunches
$A$ and $B$. $A: B$ denotes bunch inclusion --- bunch $A$ is included
in bunch $B$. We use notation $x,..y$ to mean from (including) $x$ to
(excluding) $y$.

If $x$ is a fresh (previously unused) name, $D$ is a bunch, and $b$ is
an arbitrary expression, then $\lambda x: D \cdot b$ is a
\emph{function} of a variable (parameter) $x$ with domain $D$ and body
$b$. If $f$ is a function, then $\Delta f$ denotes the domain of $f$.
If $x : \Delta f$, then $f x$ ($f$ applied to $x$) is the
corresponding element in the range. A function of $n$ variables is a
function of $1$ variable, whose body is a function of $n-1$ variables,
for $n > 0$. A predicate is function whose body is a boolean
expression. A relation is a function whose body is a predicate. A
higher-order function is a function whose parameter is a function.

A \emph{quantifier} is a unary prefix operator that applies to
functions. If $p$ is a predicate, then $\forall p$ is the boolean
result, obtained by first applying $p$ to all the elements in its
domain and then taking the conjunction of those results. Taking the
disjunction of the results produces $\exists p$. Similarly, if $f$ is
a numeric function, then $\sum f$ is the numeric result, obtained by
first applying $f$ to all the elements in its domain and then taking
the sum of those results.

\notation{$\forall \; \exists \; \sum$}{quantifiers}

For example, applying the quantifier $\sum$ to the function $\lambda
i:0,..2^n \cdot |\psi i|^2$, for some function $\psi$, yields: $\sum
\lambda i:0,..2^n \cdot |\psi i|^2$, which for the sake of simplicity
we abbreviate to $\sum i:0,..2^n \cdot |\psi i|^2$. In addition, we
allow a few other simplifications. For example, we can omit the domain
of a variable if it is clear from the context. We can also group
variables from several quantifications. For example, $\sum i:0,..2^n
\cdot \sum j:0,..2^n \cdot 2^{-m-n}$ can be abbreviated to $\sum
i,j:0,..2^n \cdot 2^{-m-n}$.

A \emph{program} is an implemented specification. For simplicity we
only take the following to be implemented: $ok$, assignment,
\textbf{if then else}, sequential composition, booleans, numbers,
bunches, and functions.

Given a specification $S$, we proceed as follows. If $S$ is a program,
there is no work to be done. If it is not, we build a program $P$,
such that $P$ refines $S$, i.e. $S \Leftarrow P$. The refinement can
proceed in steps: $S \Leftarrow \hdots \Leftarrow R \Leftarrow Q
\Leftarrow P$.

One of the best features of Hehner's theory, is its simple treatment
of recursion. In $S \Leftarrow P$ it is possible for $S$ to appear in
$P$. No additional rules are required to prove the refinement. For example,
it is trivial to prove that
\begin{equation*}
  x \geq 0 \Rightarrow x'=0 
  \lar
  \textbf{if } x=0
  \textbf{ then } ok 
  \textbf{ else } (x:=x-1; x \geq 0 \Rightarrow x'=0 )
\end{equation*}
The specification says that if the initial value of $x$ is non-negative,
its final value must be $0$. The solution is: if the value of $x$ is
zero, do nothing, otherwise decrement $x$ and repeat.

How long does the computation take? To account for time we add a time
variable $t$. We use $t$ to denote the time, at which the computation
starts, and $t'$ to denote the time, at which the computation ends. In
case of non-termination, $t'=\infty$. This is the only characteristic
by which we distinguish terminating programs from non-terminating
ones. See~\cite{hehner06} for a discussion on treatment of
termination. We choose to use a \emph{recursive time} measure, in
which we charge $1$ time unit for each time $P$ is called. We replace
each call to $P$ to include the time increment as follows:
\begin{equation*}
  P
  \lar
  \textbf{if } x=0
  \textbf{ then } ok 
  \textbf{ else } (x:=x-1; t:=t+1; P )
\end{equation*}
It is easy to see that $t$ is incremented the same number of times
that $x$ is decremented, i.e. $t' = t + x$, if $x \geq 0$, and
$t'=\infty$, otherwise. Just as above, we can prove:
\notation{$\infty$}{infinity}
\begin{align*}
  & x \geq 0 \wedge t' = t + x \vee x  < 0 \wedge t'= \infty\\
  \lar
  & \textbf{if } x=0
    \textbf{ then } ok\\ 
  & \textbf{ else } (x:=x-1; \, t:=t+1; \,
                    x \geq 0 \wedge t' = t + x \vee
                    x < 0 \wedge t'= \infty )
\end{align*}

\subsubsection{Probabilistic predicative programming}
\label{sec:sub_ppp}

Probabilistic predicative programming was introduced
in~\cite{hehner04:_book_2ed} and was further developed
in~\cite{hehner04:_ppp}. It is a generalisation of predicative
programming that allows reasoning about probability distributions of
values of variables of interest. Although in this work we apply this
reasoning to boolean and integer variables only, the theory does not
change if we want to work with real numbers: we replace summations
with integrals.

A \emph{probability} is a real number between $0$ and $1$, inclusive.
A \emph{distribution} is an expression whose value is a probability
and whose sum over all values of variables is $1$. For example, if $n$
is a positive natural variable, then $2^{-n}$ is a distribution, since
for any $n$, $2^{-n}$ is a probability, and $\sum n \cdot
2^{-n}=1$. In two positive natural variables $m$ and $n$, $2^{-n-m}$
is also a distribution. If a distribution of several variables can be
written as a product of distributions of the individual variables,
then the variables are \emph{independent}. For example, $m$ and $n$ in
the previous example are independent. Given a distribution of several
variables, we can sum out some of the variables to obtain a
distribution of the rest of the variables. In our example, $\sum n
\cdot 2^{-n-m} = 2^{-m}$, which is a distribution of $m$.

To generalise boolean specifications to probabilistic specifications,
we use $1$ and $0$ for boolean $true$ and $false$,
respectively.\footnote{Readers familiar with $\top$ and $\bot$
  notation can notice that we take the liberty to equate $\top=1$ and
  $\bot=0$.} If $S$ is an implementable deterministic specification
and $p$ is a distribution of the initial state $x, y, ...$, then the
distribution of the final state is
\begin{equation*}
  \sum {x, y, ...} \cdot S \times p
\end{equation*}
If $R$ and $S$ are specifications in variables $x, y, \hdots \;$, $R''$
is obtained from $R$ by substituting all occurrences of primed
variables $x', y', \hdots$ with double-primed variables $x'', y'',
\hdots \;$, and $S''$ is obtained from $S$ by substituting all
occurrences of unprimed variables $x, y, \hdots$ with double-primed
variables $x'', y'', \hdots \;$, then the \emph{sequential composition}
of $R$ and $S$ is defined by
\begin{equation*}
  R;S \eqv \sum {x'', y'', \hdots} \cdot R'' \times S''
\end{equation*}
%
\notation{$P;R$}{sequential composition (generalised)}
If $p$ is a probability and $R$ and $S$ are distributions, then
\begin{equation*}
  \textbf{if } p \textbf{ then } R \textbf{ else } S \eqv
  p \times R + (1-p)\times S
\end{equation*}
\notation{$\textbf{if then else}$}{generalised}
Various laws can be proven about sequential composition. One of the
most important ones, the substitution law, introduced earlier, applies
to probabilistic specifications as well.

To implement a probabilistic specification we use a pseudo-random
number generator. Since we cannot, even in theory, produce a real
random number generator by means of traditional computing, we assume
that a pseudo-random number generator generates truly random numbers
and we simply refer to it as random number generator. For a positive
natural variable $n$, we say that $rand \; n$ produces a random
natural number uniformly distributed in $0,..n$. To reason about the
values supplied by the random number generator consistently, we replace
every occurrence of $rand \; n$ with a fresh variable $r$ whose value
has probability $(r: 0,..n)/n$. If $rand$ occurs in a context such as
$r = rand \; n$, we replace the equation by $r:(0,..n)/n$. If $rand$
occurs in the context of a loop, we parametrise the introduced
variables by the execution time.

Recall the earlier example. Let us change the program slightly by
introducing probabilism:
\begin{equation*}
  P  
  \lar
  \textbf{if } x=0
  \textbf{ then } ok 
  \textbf{ else } (x:= x - rand\; 2; t:=t+1; P )
\end{equation*}
In the new program at each iteration $x$ is either decremented by $1$
or it is unchanged, with equal probability.  Our intuition tells us
that the revised program should still work, except it should take
longer. Let us prove it. We replace $rand$ with $r:time
\rightarrow (0,1)$ with $rt$ having probability $1/2$. Ignoring time
we can prove:
\begin{align*}
  & x \geq 0 \Rightarrow x'=0\\
  \lar
  & \textbf{if } x=0
    \textbf{ then } ok  
    \textbf{ else } (x:=x - rand\; 2; x \geq 0 \Rightarrow x'=0 )
\end{align*}
As for the execution time, we can prove that it takes at least $x$
time units to complete:
\begin{alignat*}{2}
  & t' \geq t+x\\
  \lar
  &  \textbf{if } x=0
    \textbf{ then } ok  
    \textbf{ else } (x:=x - rand\; 2; t:=t+1; t' \geq t + x )
\end{alignat*}
How long should we expect to wait for the execution to complete? In
other words, what is the distribution of $t'$? Consider the following
distribution of the final states:
\begin{align*}
  &  (0 = x' = x =  t'-t )  +
     (0 = x' < x \leq t'-t) \times
     \binom{t'-t-1}{x-1} \times \frac{1}{2^{t'-t}}, \\
  & \text{ where } \binom{n}{m} = \frac{n!}{m! \times (n-m)!}
\end{align*}
We can prove that:
\begin{align*}
 &\sum {rt} \cdot
    \frac{1}{2} \times
    \left( \vphantom{\binom{t'-t-1}{x-1}} \right.
   \begin{aligned}[t]
      & \textbf{if } x=0
        \textbf{ then } ok   \\
      & \textbf{else }
        \left( \vphantom{\binom{t'-t-1}{x-1}} \right.
          \begin{aligned}[t]
            & x:=x - rt; \; t:=t+1; \\
            & (0 = x' = x =  t'-t ) \; + \\
            & (0 = x' < x \leq t'-t) \;\times 
              \left. \left.
               \binom{t'-t-1}{x-1} \times
               \frac{1}{2^{t'-t}} \right) \right)
          \end{aligned}
    \end{aligned} \br
\eqv
 & (0 = x' = x = t'=t) \; + 
   (0 = x' < x \leq t'-t) \times
    \binom{t'-t-1}{x-1} \times
    \frac{1}{2^{t'-t}}
\end{align*}
Now, since for positive $x$, $t'$ is distributed according to the
negative binomial distribution with parameters $x$ and $\frac{1}{2}$,
its mean value is
\begin{align*}
 &\hspace{-2.5mm}\sum {t'} \cdot (t'-t)\times
   \left((0 = x =  t'-t )  +
    (0 < x \leq t'-t) \times \binom{t'-t-1}{x-1} \times
    \frac{1}{2^{t'-t}}\right) \\
 &\hspace{-2mm}\eqv
  2 \times x + t
\end{align*}
Therefore, we should expect to wait $2\times x$ time units for the
computation to complete.



\section{Quantum Predicative Programming}
\label{sec:qpp}

This section is the contribution of the paper. Here we define the
quantum system, introduce programming with the quantum system and
several well-known problems, their classical and quantum solutions,
and their formal comparative time complexity analyses. The proofs of
refinements are omitted for the sake of brevity. The reader is
referred to~\cite{tafliovich04} for detailed proofs of some of the
algorithms.

\subsection {The quantum system}
\label{sec:qsystem}

Let $\mathbb{C}$ be the set of all complex numbers with the absolute
value operator $|\cdot|$ and the complex conjugate operator $^*$. Then
a state of an $n$-qubit system is a function $\psi : 0,..2^n
\rightarrow \mathbb{C}$, such that $\sum {x:0,..2^n} \cdot |\psi x|^2
= 1$\footnote{We should point out that this kind of function
  operations is referred to as \emph{lifting}}.

If $\psi$ and $\phi$ are two states of an $n$-qubit system, then their
\emph{inner product}, $\braket{\psi}{\phi} : \mathbb{C}$, is defined by: 
\begin{equation*}
  \braket{\psi}{\phi} = \sum {x:0,..2^n} \cdot (\psi x)^* \times (\phi x)
\end{equation*}
A \emph{basis} of an $n$-qubit system is a collection of $2^n$ quantum
states $b_{0,..2^n}$, such that $\forall i,j:0,..2^n \cdot
\braket{b_i}{b_j} = (i=j)$.

We adopt the following Dirac-like notation for the computational
basis: if $x: 0,..2^n$, then $\textbf{x}$ denotes the corresponding
$n$-bit binary encoding of $x$ and $\ketb{x}: 0,..2^n \rightarrow
\mathbb{C}$ is the following quantum state:
\begin{equation*}
  \ketb{x} = \lambda i:0,..2^n \cdot (i=x)
\end{equation*}
If $\psi$ is a state of an $m$-qubit system and $\phi$ is a state of
an $n$-qubit system, then $\psi \tensor \phi$, the tensor product of
$\psi$ and $\phi$, is the following state of a composite $m+n$-qubit
system:
\begin{equation*}
  \psi \tensor \phi = \lambda i:0,..2^{m+n} \cdot 
                      \psi (i\:div\:2^n) \times \phi (i\:mod\:2^n)
\end{equation*}
We write $^{\tensor n}$ to mean \emph{tensored with itself $n$ times}.

An operation defined on a $n$-qubit quantum system is a higher-order
function, whose domain and range are maps from $0,..2^n$ to the
complex numbers. An \emph{identity} operation on a state of an
$n$-qubit system is defined by
\begin{equation*}
  I^n = \lambda \psi : 0,..2^n \rightarrow \mathbb{C} \cdot \psi
\end{equation*}
For a linear operation $A$, the \emph{adjoint} of $A$, written
$A^\dagger$, is the (unique) operation, such that for any two states
$\psi$ and $\phi$, $\braket{\psi}{A \phi} = \braket{A^{\dagger}
  \psi}{\phi}$.
The \emph{unitary transformations} that describe the evolution of a
$n$-qubit quantum system are operations $U$ defined on the system,
such that $U^{\dagger} U = I^n$.
In this setting, the \emph{tensor product} of operators is defined in
the usual way. If $\psi$ is a state of an $m$-qubit system, $\phi$ is
a state of an $n$-qubit system, and $U$ and $V$ are operations defined
on $m$ and $n$-qubit systems, respectively, then the tensor product of
$U$ and $V$ is defined on an $m+n$ qubit system by $(U \tensor V)
(\psi \tensor \phi) = (U \psi) \tensor (V \phi)$.

Just as with tensor products of states, we write $U^{\tensor n}$ to mean
\emph{operation $U$ tensored with itself $n$ times}. 

Suppose we have a system of $n$ qubits in state $\psi$ and we measure
it. Suppose also that we have a variable $r$ from the domain
$0,..2^n$, which we use to record the result of the measurement, and
variables $x,y, \hdots$, which are not affected by the measurement.
Then the measurement corresponds to a probabilistic specification that
gives the probability distribution of $\psi'$ and $r'$ (these depend
on $\psi$ and on the type of measurement) and states that the
variables $x,y,\hdots$ are unchanged.

For a general quantum measurement described by a collection $M =
M_{0,..2^n}$ of measurement operators, which satisfy the completeness
equation $\sum {m:0,..2^n} \cdot M_m^\dagger M_m = I$, the
specification is $\textbf{measure}_M \, \psi \, r$, where
\begin{equation*}
  \textbf{measure}_M \, \psi \, r \eqv
    \braket{\psi}{M_{r'}^\dagger M_{r'}\psi} \times
    \left(
     \psi'=\frac {M_{r'}\psi}{\sqrt{\braket{\psi} {M_{r'}^\dagger M_{r'}\psi}}}
    \right) \times (\sigma'=\sigma)
\end{equation*}
where $\sigma'=\sigma$ is an abbreviation of $(x'=x) \times (y'=y)
\times \hdots$ and means ``all other other variables are unchanged''.
\notation{\textbf{measure}}{quantum measurement}
To obtain the distribution of, say, $r'$ we sum out the rest of the
variables as follows:
\begin{align*}
 &\sum {\psi', x', y', \hdots} \cdot
    \braket{\psi}{M_{r'}^\dagger M_{r'}\psi} \times \!
    \left(
     \psi'=\frac {M_{r'}\psi}{\sqrt{\braket{\psi} {M_{r'}^\dagger M_{r'}\psi}}}
    \right)\! \times\! (\sigma'=\sigma)\\
\eqv
 &\braket{\psi}{M_{r'}^\dagger M_{r'}\psi}
\end{align*}
For projective measurements defined by an observable $O = \sum m
\cdot \lambda_m \times P_m$, where $P_m$ is the projector on the
eigenspace of $O$ with eigenvalue $\lambda_m$:
\begin{equation*}
  \textbf{measure}_O \, \psi \, r \eqv
    \braket{\psi}{P_{r'} \psi} \times
     \left( \psi'=\frac {P_r' \psi}{\sqrt{\braket{\psi}{P_r' \psi}}}
     \right) \times (\sigma'=\sigma)
\end{equation*}
Given an arbitrary orthonormal basis $B = b_{0,..2^n}$, measurement of
$\psi$ in basis $B$ is:
\begin{equation*}
  \textbf{measure}_B \, \psi \, r \eqv 
    |\braket{b_{r'}}{\psi}|^2 \times (\psi'=b_{r'}) \times
      (\sigma'=\sigma)
\end{equation*}
Finally, the simplest and the most commonly used measurement in the
computational basis is:
\begin{equation*}
  \textbf{measure } \psi \, r \eqv
    | \psi r' |^2  \times (\psi'= \ketb{r'}) \times (\sigma'=\sigma)
\end{equation*}
In this case the distribution of $r'$ is $|\psi r'|^2$ and the
distribution of the quantum state is:
\begin{equation*}
  \sum {r'} \cdot |\psi r'|^2 \times (\psi' = \ketb{r'})
\end{equation*}
which is precisely the mixed quantum state that results from the
measurement. 

In order to develop quantum programs we need to add to our list of
implemented things from section~\ref{sec:ppp}. We add variables of
type quantum state as above and we allow the following three kinds of
operations on these variables.  If $\psi$ is a state of an $n$-qubit
quantum system, $r$ is a natural variable, and $M$ is a collection of
measurement operators that satisfy the completeness equation, then:
\begin{enumerate}
\item $\psi:=\ket{0}^{\tensor n}$ is a program
\item $\psi:=U\psi$, where $U$ is a unitary transformation on an $n$-qubit system,
      is a program
\item $\textbf{measure}_M \, \psi \, r$ is a program
\end{enumerate}
The special cases of measurements, described in section~\ref{sec:qc},
are therefore also allowed: for an observable $O$, $\textbf{measure}_O
\, q\,r$ is a program; for an orthonormal basis $B$,
$\textbf{measure}_B \, q\,r$ is a program; finally, $\textbf{measure }
q\,r$ is a program.

\subsection{Deutsch-Jozsa algorithm}
\label{sec:deutchjosza}

Deutsch-Jozsa problem (\cite{jozsa91}), an extension of Deutsch's
Problem (\cite{deutsch85}), is an example of the broad class of
quantum algorithms that are based on quantum Fourier transform
(\cite{jozsa98}).  The task is: given a function $f:0,..2^n
\rightarrow 0,1$ , such that $f$ is either constant or balanced,
determine which case it is.  Without any restrictions on the number of
calls to $f$, we can write the specification (let us call it $S$) as
follows:
\begin{equation*}
  (f \; is \; constant \vee f \; is \; balanced) \rar b'= f \; is \; constant
\end{equation*}
where $b$ is a boolean variable and the informally stated properties of
$f$ are defined formally as follows:
\begin{align*}
 & f \; is \; constant \eqv \forall {i:0,..2^n} \cdot fi=f0  \\
 & f \; is \; balanced \eqv 
                 \left | \sum {i:0,..2^n}\cdot (-1)^{fi} \right| = 0
\end{align*}
It is easy to show that
\begin{align*}
  & (f \; is \; constant \vee f \; is \; balanced) \\
  \rar
  &  (f \; is \; constant \eqv \forall {i:0,..2^{n-1}+1} \cdot fi=f0)
\end{align*}
In our setting, we need to implement the specification $R$ defined as follows:
\begin{equation*}
  b' \eqv \forall {i:0,..2^{n-1}+1} \cdot fi=f0
\end{equation*}
The \emph{Hadamard} transform, widely used in quantum algorithms, is
defined on a $1$-qubit system and in our setting is a higher-order
function from $0,1 \rightarrow \mathbb{C}$ to $0,1 \rightarrow
\mathbb{C}$:
\begin{equation*}
  \label{eq:hadamard}
  H = \lambda \psi:0,1 \rightarrow \mathbb{C} \cdot i:0,1 \cdot
       (\psi 0 + (-1)^i \times  \psi 1)/\sqrt2
\end{equation*}
\notation{$H$}{Hadamard transform on states}
The operation $H^{\tensor n}$ on a $n$-qubit system applies $H$ to
every qubit of the system. Its action on a zero state of an $n$-qubit
system is:
\begin{equation*}
  \label{eq:hadamard_on_0}
  H^{\tensor n} \ket{0}^{\tensor n} =
     \sum {x:0,..2^n} \cdot \ketb{x} / \sqrt{2^n}
\end{equation*}
On a general state $\ketb{x}$, the action of $H^{\tensor n}$ is:
\begin{equation*}
  \label{eq:hadamard_on_x}
  H^{\tensor n} \ketb{x} =
    \sum {y:0,..2^n} \cdot 
      (-1)^{\textbf{x} \cdot \textbf{y}} \times \ketb{y} / \sqrt{2^n}
\end{equation*}
where $\textbf{x} \cdot \textbf{y}$ is the bitwise inner product of 
$\textbf{x}$ and $\textbf{y}$ modulo 2 (bitwise XOR).

Another important definition is that of the quantum analog of a
classical oracle $f$:
\begin{equation*}
  U_f = \lambda \psi:0,1 \rightarrow \mathbb{C} \cdot x:0,1 \cdot
         (-1)^{fx} \times \psi x
\end{equation*}

The quantum solution, in one quantum variable $\psi$ and an integer
variable $r$ is:
\begin{equation*}
  \psi:=\ket{0}^{\tensor n} ;\,
  \psi:=H^{\tensor n}\psi ;\,
  \psi:=U_f\psi ;\,
  \psi:=H^{\tensor n}\psi ;\,
  \textbf{measure } \psi\,r  ;\,
  b:=(r'=0)
\end{equation*}
Let us add to the specification a restriction on the number of calls
to the oracle by introducing a time variable. Suppose the new
specification is:
\begin{equation*}
 (f \; is \; constant \vee f \; is \; balanced \rar b'= f \; is \; constant)
 \wedge (t'=t+1)
\end{equation*}
where we charge 1 unit of time for each call to the oracle and all
other operations are free.  Clearly, the above quantum solution works.
Classically the specification is unimplementable. In fact, the
strongest classically implementable specification is
\begin{equation*}
  (f \; is \; constant \vee f \; is \; balanced \rar b'= f \; is \; constant)
 \wedge (t'=t+2^{n-1}+1)
\end{equation*}

\subsection{Grover's search}
\label{sec:grover}

Grover's quantum search algorithm (\cite{grover96}) is well-known for
the quadratic speed-up it offers in the solutions of NP-complete
problems. The algorithm is optimal up to a multiplicative constant
(\cite{boyer_brassard_hoyer_tapp98}). The task is: given a function
$f:0,..2^n \rightarrow 0,1$, find $x:0,..2^n$, such that $fx=1$. For
simplicity we assume that there is only a single solution, which we
denote $x_1$, i.e. $f \, x_1 = 1$ and $f\, x = 0$ for all $x \neq
x_1$. The proofs are not very different for a general case of more
than one solutions.

As before, we use a general quantum oracle, defined by
\begin{equation*}
  U_f \ketb{x} = (-1)^{fx} \times \ketb{x}
\end{equation*}
\notation{$U_f$}{quantum oracle}
In addition, we define the \emph{inversion about mean} operator as
follows:
\begin{align*}
  &M: (0,..N \rightarrow \mathbb{C}) \rightarrow (0,..N \rightarrow \mathbb{C}) \\
  &M \psi \eqv \; \lambda x:0,..N \cdot 
      2 \times \left( \sum {i:0,..N} \cdot \psi i / N \right) - \psi x
\end{align*}
where $N=2^n$.

\notation{$M$}{inversion about mean}

Grover's algorithm initialises the quantum system to an equally
weighted superposition of all basis states  $\ketb{x}, \, x:0,..N$.
It then repeatedly applies $U_f$ followed by $M$ to the system.
Finally, the state is measured. The probability of error is determined
by the number of iterations performed by the algorithm. 

The algorithm is easily understood with the help of a geometric
analysis of the operators.  Let $\alpha$ be the sum over all $x$,
which are not solutions, and let $\beta$ be the solution:
\begin{align*}
  &\alpha = \frac{1}{\sqrt{N-1}} \times \sum {x \neq x_1} \cdot \ketb{x} \\
  &\beta  = \ketb{x$_1$}
\end{align*}
Then the oracle $U_f$ performs a \emph{reflection} about the vector
$\alpha$ in the plane defined by $\alpha$ and $\beta$. In other words,
$U_f(a \times \alpha + b \times \beta) = a \times \alpha - b \times
\beta$. Similarly, the inversion about mean operator is a reflection
about the vector $\psi$ in the plane defined by $\alpha$ and $\beta$.
Therefore, the result of $U_f$ followed by $M$ is a \emph{rotation} in
this plane.
The quantum solution, in a quantum variable $\psi$, natural variables
$r$, $i$, and $k$, and time variable $t$, is,
\begin{align*}
 S
\eqv
  &\left(
     \sin \left(
            (2\times(t'-t)+1) \times \arcsin \sqrt{1/N}
          \right)
   \right)^2 \;
   \times (r'=x_1) \; +\\
  &\left(1 - \left(
               \sin \left(
                      (2\times(t'-t)+1)\times \arcsin \sqrt{1/N}
                    \right)
             \right)^2
   \right) 
   \times \\
  &\;\;(r' \neq x_1) / (N-1) \br
\eqv
  & P ; \; \textbf{measure } \psi \; r \br
 P 
\lar
  & i:=0; \; \psi:=\ket{0}^{\tensor n}; \; \psi:=H^{\tensor n} \psi; \; R \br
 R
\lar
  & \textbf{if } i=k \textbf{ then } ok\\
  &  \textbf{else } 
      (i:=i+1; \; t:=t+1; \; \psi:=U_f \psi; \; \psi:=M \psi; \; R)
\end{align*}
Specification $S$ carries a lot of useful information. For example, it
tells us that the probability of finding a solution after $k$
iterations is 

$$\left(\sin ((2\times k+1) \times \arcsin \sqrt{1/N})\right)^2$$

Or we might ask how many iterations should be performed to minimise
the probability of an error. Examining first and second derivatives,
we find that the above probability is minimised when $t'-t = (\pi
\times i)/(4 \times \arcsin \sqrt{1/N}) - 1/2$ for integer $i$. Of
course, the number of iterations performed must be a natural number.
It is interesting to note that probability of error is periodic in the
number of iterations, but since we don't gain anything by performing
extra iterations, we pick $i=1$. Finally, assuming $1 \ll N = 2^n$, we
obtain an elegant approximation to the optimal number of iterations:
$\left \lceil \pi \times \sqrt{2^n} / 4 \right \rceil$, with the
probability of error approximately $1/2^n$.

\subsection {Computing with Mixed States}
\label{sec:computing_mixed}

As we have discussed in section~\ref{sec:qc}, the state of a quantum
system after a measurement is traditionally described as a \emph{mixed
  state}. An equation $\psi = \{ \ket{0} \} /2 + \{ \ket{1} \} /2$
should be understood as follows: the state $\psi$ is $\ket{0}$ with
probability $1/2$ and it is $\ket{1}$ with probability $1/2$. In
contrast to a pure state, a mixed state does not describe a physical
state of the system.  Rather, it describes our knowledge of in what
state the system is.

In our framework, there is no need for an additional mechanism to
compute with mixed states. Indeed, a mixed state is not a system
state, but a distribution over system states, and all our programming
notions apply to distributions. The above mixed state is the following
distribution over a quantum state $\psi$: $(\psi = \ket{0}) /2 + (\psi
= \ket{1}) / 2$. This expression tells us, for each possible value in
the domain of $\psi$, the probability of $\psi$ having that value. For
example, $\psi$ is the state $\ket{0}$ with probability $(\ket{0} =
\ket{0}) /2 + (\ket{0} = \ket{1}) / 2$, which is $1/2$; it is
$\ket{1}$ with probability $(\ket{1} = \ket{0}) /2 + (\ket{1} =
\ket{1}) / 2$, which is also $1/2$; for any scalars $\alpha$ and
$\beta$, not equal to $0$ or $1$, $\psi$ is $\alpha \times \ket{0} +
\beta \times \ket{1}$ with probability $(\alpha \times \ket{0} + \beta
\times \ket{1} = \ket{0}) /2 + (\alpha \times \ket{0} + \beta \times
\ket{1} = \ket{1}) / 2$, which is $0$.  One way to obtain this
distribution is to measure an equally weighted superposition of
$\ket{0}$ and $\ket{1}$:
\begin{alignat*}{2}
 & \psi' = \ket{0} /\sqrt2 + \ket{1} /\sqrt2; \,
   \textbf{measure } \psi \, r &\\
 && \text{measure}\br
\eqv
 & \psi' = \ket{0} /\sqrt2 + \ket{1} /\sqrt2; \,
   |\psi r'|^2 \times (\psi' = \ketb{r'}) &\\
   &&\hspace{-2cm}\text{sequential composition}\br
\eqv
 & \sum {r'', \psi''} \cdot
      (\psi'' = \ket{0} /\sqrt2 + \ket{1} /\sqrt2) 
      \times
      |\psi'' r'|^2 \times (\psi' = \ketb{r'})&\\
 &&\hspace{-2cm}\text{one point law} \br
\eqv
 & |(\ket{0} /\sqrt2 + \ket{1} /\sqrt2) \, r'|^2 \times (\psi' = \ketb{r'}) &\\
\eqv
 & (\psi' = \ketb{r'}) / 2
\end{alignat*}
Distribution of the quantum state is then:
\begin{equation*}
 \sum {r'} \cdot (\psi' = \ketb{r'}) / 2
 \eqv
 (\psi' = \ket{0}) / 2 + (\psi' = \ket{1}) / 2
\end{equation*}
as desired.

Similarly, there is no need to extend the application of unitary
operators. Consider the following toy program:
\begin{equation*}
  \psi:=\ket{0} ;\,
  \psi:=H \psi ;\,
  \textbf{measure } \psi \, r ;\,
  \textbf{if } r=0 \textbf{ then } \psi:=H \psi \textbf{ else } ok
\end{equation*}
In the second application of Hadamard the quantum state is mixed, but
this is not evident from the syntax of the program. It is only in the
analysis of the final quantum state that the notion of a mixed state
is meaningful. The operator is applied to a (pure) system state, though
we are unsure what that state is.
\begin{alignat*}{2}
  & \psi:=\ket{0} ;\,
    \psi:=H \psi ;\,
    \textbf{measure } \psi \, r \,;
   & \\
  & \textbf{if } r=0 \textbf{ then } \psi:=H \psi \textbf{ else } ok
   &\text{as before} \br
\eqv
  & (\psi' = \ketb{r'}) / 2 ;
   &\\
  & \textbf{if } r=0 \textbf{ then } \psi:=H \psi \textbf{ else } ok
   &\hspace{-1cm}\text{sequential composition} \br
\eqv
  & \sum {r'', \psi''} \cdot
     \begin{aligned}[t]
      &  (\psi'' = \ketb{r''}) / 2 \; \times \\
      & \left((r''=0) \times (\psi' = H \psi'') \times (r'=r'') \, + \right. \\
      & \left.(r''=1) \times (\psi' = \psi'') \times (r'=r'') \right)
     \end{aligned}
  & \begin{aligned}[t]
      &\\
      &\\
      &\text{one point law}
    \end{aligned} \br
\eqv
  & \left( (\psi' = H \ket{0}) \times (r'=0) +
           (\psi' = \ket{1}) \times (r'=1) \right)/2 
  &\br
\eqv
   & (\psi' = \ket{0} / \sqrt2 + \ket{1} / \sqrt2) \times (r'=0) / 2 \, + &\\
  & (\psi' = \ket{1}) \times (r'=1) /2
\end{alignat*}
The distribution of the quantum state after the computation is:
\begin{align*}
 &\sum {r'} \cdot 
    (\psi' = \ket{0} / \sqrt2 + \ket{1} / \sqrt2) \times (r'=0) / 2 +
    (\psi' = \ket{1}) \times (r'=1) /2 \\
\eqv
 &  (\psi' = \ket{0} / \sqrt2 + \ket{1} / \sqrt2) / 2 +
    (\psi' = \ket{1}) /2
\end{align*}
A lot of properties of measurements and mixed states can be proven
from the definitions of measurement and sequential composition. For
example, the fact that a measurement in the computational basis,
performed immediately following a measurement in the same basis, does
not change the state of the system and yields the same result as the
first measurement with probability $1$, is proven as follows:
\begin{alignat*}{2}
  & \textbf{measure } \psi \,r ;\,
    \textbf{measure } \psi \,r
   &\text{measure} \br
\eqv
  & |\psi \, r'|^2 \times (\psi' = \ketb{r'}) ;\,
    |\psi \, r'|^2 \times (\psi' = \ketb{r'})
   &\hspace{-1.1cm}\text{sequential composition} \br
\eqv
  &\sum {\psi'', r''} \cdot
    |\psi \, r''|^2 \times (\psi'' = \ketb{r''}) \times
    |\psi'' \, r'|^2 \times (\psi' = \ketb{r'})
   &\hspace{-1cm}\text{one point law} \br
\eqv
  & |\psi \, r'|^2 \times (\psi' = \ketb{r'})
   &\text{measure} \br
\eqv
  & \textbf{measure } \psi \,r
\end{alignat*}
In case of a general quantum measurement, the proof is similar, but a
little more computationally involved.



\section {Conclusion and Future Work}
\label{sec:conclusion}

We have presented a new approach to developing, analysing, and proving
correctness of quantum programs. Since we adopt Hehner's theory as the
basis for our work, we inherit its advantageous features, such as
simplicity, generality, and elegance. Our work extends probabilistic
predicative programming in the same fashion that quantum computation
extends probabilistic computation. We have provided tools to write
quantum as well as classical specifications, develop quantum and
classical solutions for them, and analyse various properties of
quantum specifications and quantum programs, such as implementability,
time and space complexity, and probabilistic error analysis uniformly,
all in the same framework.

Current research an research in the immediate future involve reasoning
about distributed quantum computation. Current work involves
expressing quantum teleportation, dense coding, and various games
involving entanglement, in a way that makes complexity analysis of
these quantum algorithms simple and natural. These issues will be
described in a forthcoming paper. We can easily express teleportation
as refinement of a specification $\phi' = \psi$, for distinct qubits
$\phi$ and $\psi$, in a well-known fashion. However, we are more
interested in the possibilities of simple proofs and analysis of
programs involving communication, both via quantum channels and
exhibiting the LOCC (local operations, classical communication)
paradigm. Future work involves formalising quantum cryptographic
protocols, such as BB84~\cite{bennet84}, in our framework and provide
formal analysis of these protocols. This will naturally lead to formal
analysis of distributed quantum algorithms (e.g. distributed Shor's
algorithm of~\cite{yimsiriwattana04}).



\bibliographystyle{plain}
\bibliography{qpp}

\end{document}